\documentclass[sort&compress,final]{aipproc}
\layoutstyle{8x11single}
\usepackage{amsmath}
\usepackage{amssymb}
\newcommand{\be}{\begin{equation}}
\newcommand{\ee}{\end{equation}}
\newcommand{\MS}{\overline{\mathrm{MS}}}
\newcommand{\MC}{\mathrm{MC}}
\newcommand{\NP}{\mathrm{NP}}
\newcommand{\G}{\mathrm{G}}
\newcommand{\latt}{\mathrm{latt}}
\newcommand{\nn}{\nonumber}
\newcommand{\lQ}{\Lambda_{\mathrm{QCD}}}
\newcommand{\al}{\alpha}

\newcommand{\Eqre}{Eq.~\eqref}

\begin{document}

\title{Phenomenology of renormalons and the OPE from lattice
regularization: 
the gluon condensate and the heavy quark pole mass}


\author{Gunnar S.\ Bali}{
  address={Institut f\"ur Theoretische Physik, Universit\"at
Regensburg, D-93040 Regensburg, Germany} ,
altaddress={Tata Institute of Fundamental Research, Homi Bhabha Road, Mumbai 400005, India}
}
\author{Antonio Pineda}{
  address={Grup de F\'{\i}sica Te\`orica and IFAE, Universitat
Aut\`onoma de Barcelona, E-08193 Bellaterra, Barcelona, Spain}
}
\begin{abstract}
 We study the operator product expansion of the plaquette (gluon condensate) and the self-energy of an infinitely heavy quark. We first 
 compute their perturbative expansions to order $\alpha^{35}$ and $\alpha^{20}$, respectively, in the lattice scheme. In both cases we reach the asymptotic regime  where the renormalon behavior sets in. Subtracting the perturbative series, we obtain 
 the leading non-perturbative corrections of their respective operator product expansions. In the first case we obtain the gluon condensate and in the second the binding energy of the heavy quark in the infinite mass limit. The results are fully consistent with the expectations from renormalons and the operator product expansion.
\end{abstract}

\maketitle

\section{Introduction}

The operator product expansion (OPE)~\cite{Wilson:1969zs} is a fundamental
tool for theoretical analyses in quantum field theories.
Its validity is only proven
rigorously within perturbation theory, to arbitrary finite
orders~\cite{Zimmermann:1972tv}. The use of the OPE
in a non-perturbative framework was initiated
by the ITEP group~\cite{Vainshtein:1978wd}
(see also the discussion in~Ref.~\cite{Novikov:1984rf}), who postulated
that the OPE of a correlator could be approximated by the following series:
\be
\label{eq:ope1}
\mathrm{correlator}(Q) \simeq \sum_d\frac{1}{Q^d}C_d(\alpha)
\langle O_d \rangle
\,,
\ee  
where the expectation values of local operators $O_d$
are suppressed by inverse powers of a large
external momentum $Q\gg\lQ$, according to their dimensionality $d$.
The Wilson coefficients $C_d(\alpha)$ 
encode the physics at momentum scales larger than $Q$.
These are well approximated by perturbative
expansions in the strong coupling parameter $\alpha$:
\be
C_d(\alpha) \simeq \sum_{n\geq 0} c_n\al^{n+1}\,.
\ee
The large-distance
physics is described by the matrix elements
$\langle O_d \rangle$ that usually have to be
determined non-perturbatively: $\langle O_d \rangle \sim \lQ^d$.

It can hardly be overemphasized that (except for direct
predictions of non-perturbative lattice simulations, e.g., on
light hadron masses)
all QCD predictions are based on factorizations that are generalizations of the above generic OPE.

There exist some major questions related to the OPE
that have to be addressed:\footnote{There are also some important points
that we do not address here, including:\\
We do not consider ambiguities associated to short-distance
non-perturbative effects, which would give rise to
singularities further away from the origin of the Borel plane 
than those we study here.\\
We take the validity of the OPE in pure perturbation theory for granted.
This assumption is solid in cases with a single large scale, $Q^2$, and
in Euclidean spacetime.\\
We will not discuss the validity of the (non-perturbative) OPE
for timelike distances that can occur in Minkowski
spacetime, an issue related to possible violations of
quark-hadron duality.}
\begin{itemize}
\item
Are the perturbative expansions of Wilson coefficients asymptotic series?
\item
If so: are the associated ambiguities of the asymptotic behavior
consistent with the OPE, i.e.\ with the positions of the
expected renormalons~\cite{Hooft} in the Borel plane?
\item
Is the OPE valid beyond perturbation theory?
\item
What is the real size of the first non-perturbative
correction within a given OPE expansion?
\item
Is this value strongly affected by ambiguities associated to renormalons?
\end{itemize}
In this paper we summarize and discuss our
recent results~\cite{Bauer:2011ws,Bali:2013pla,Bali:2013qla,Bali:2014fea,Bali:2014sja}, which 
address these questions for the case of the plaquette and the energy of an infinitely heavy quark in the pure gluodynamics approximation to QCD.
Both analyses utilize lattice regularization. Contrary to, e.g., dimensional regulation, lattice regularization
can be defined non-perturbatively. Using a lattice scheme rather than the $\MS$ scheme, we can, not only expand observables in perturbation theory, but also evaluate them non-perturbatively. Another advantage of this choice is that it enables
us to use numerical stochastic perturbation
theory~\cite{DRMMOLatt94,DRMMO94,DR0} to obtain perturbative expansion
coefficients. This allows us to realize much higher orders than
would have been possible with diagrammatic techniques.
A disadvantage of the lattice scheme is that, at least in our discretization,
lattice perturbative expansions converge slower than expansions in
the $\MS$ coupling. This means that we have to go to comparatively higher
orders to become sensitive to the asymptotic behavior. Many of the results
obtained in a lattice scheme either directly apply to the $\MS$
scheme too or can subsequently easily (and in some cases exactly)
be converted into
this scheme.

In our studies we used the Wilson gauge action~\cite{Wilson:1974sk}. 
We define the vacuum expectation value of a generic operator $B$ of
engineering dimension zero as
\begin{equation}
\label{eq:ANP}
\langle B\rangle \equiv \langle \Omega |B|\Omega\rangle =\frac{1}{Z}\int\![dU_{x,\mu}]\,e^{-S[U]} B[U]
\end{equation}
with the partition function $Z=\int[dU_{x,\mu}]\,e^{-S[U]}$
and measure $[dU_{x,\mu}]=\prod_{x\in\Lambda_E,\mu}dU_{x,\mu}$.
$|\Omega\rangle$ denotes the vacuum state,
$\Lambda_E$ is a Euclidean spacetime lattice with lattice spacing
$a$ and $U_{x,\mu}\approx e^{iA_{\mu}(x+a/2)}\in\mathrm{SU}(3)$ is a gauge link.

\section{The Plaquette: OPE in perturbation theory}

For the case of the plaquette we have $B \rightarrow P$, where
\begin{equation}
\langle P\rangle=\frac{1}{N^4}\sum_{x\in\Lambda_E}\langle P_x\rangle\,,\quad
P_{x}=1-\frac{1}{36}
\sum_{\mu>\nu}\mathrm{Tr}\left(U_{x,\mu\nu}+U_{x,\mu\nu}^{\dagger}\right)\,,
\end{equation}
and $U_{x,\mu\nu}$ denotes the oriented product of gauge links enclosing an
elementary square (plaquette) in the $\mu$-$\nu$ plane of the lattice.
For details on
the notation and simulation set-up see Ref.~\cite{Bali:2014fea}.

$\langle P\rangle$ will depend on the lattice extent $Na$, the spacing $a$ and
$\al =g^2/(4\pi) \equiv \al(a^{-1})$ (note that $\al$ is the bare lattice coupling 
and its natural scale is of order $a^{-1}$). We first compute this expectation value in strict perturbation theory. In other words, we Taylor expand in powers of $g$ {\it before} averaging over the gauge configurations (which we do using NSPT~\cite{DRMMOLatt94,DRMMO94,DR0}). 
The outcome is a power series in $\al$:
\begin{equation}
\nn
\langle P \rangle_{\mathrm{pert}}(N) \equiv \frac{1}{Z}\left.\int\![dU_{x,\mu}]\,e^{-S[U]} P[U]\right|_{\mathrm{NSPT}}
=\sum_{n\geq 0}p_n(N)\al^{n+1}\,.
\end{equation}
The dimensionless coefficients $p_n(N)$ are functions of the linear
lattice size $N$. We emphasize that they
do not depend on the lattice spacing $a$ or on the physical
lattice extent $Na$ alone but only on
the ratio $N=(Na)/a$.

We are interested in the large-$N$ (i.e.\ infinite volume)
limit. In this situation 
\be
\label{eq:scales}
\frac{1}{a} \gg \frac{1}{Na}
\ee
and it makes sense to factorize the contributions of the
different scales within the OPE
framework. 
The hard modes, of scale $\sim 1/a$,
determine the Wilson
coefficients, whereas the soft modes, of scale $\sim 1/(Na)$, can be described
by expectation values of local gauge invariant operators. There are no
such operators of dimension two.
The renormalization group invariant definition of the gluon condensate
\be
\label{eq:GC}
\langle G^2 \rangle=-\frac{2}{\beta_0}\left\langle\Omega\left| \frac{\beta(\alpha)}{\alpha}
G_{\mu\nu}^cG_{\mu\nu}^c\right|\Omega\right\rangle
=
\left\langle\Omega\left|  \left[1+\mathcal{O}(\alpha)\right]\frac{\alpha}{\pi}
G_{\mu\nu}^cG_{\mu\nu}^c\right|\Omega\right\rangle
\ee
is the only local gauge invariant expectation value of an operator of dimension
$a^{-4}$ in pure gluodynamics. In the purely perturbative case discussed here,
this only depends on the soft scale $1/(Na)$, i.e.\ on the lattice
extent. 
On dimensional grounds, the perturbative gluon condensate
$\langle G^2 \rangle_{\mathrm{soft}}$ is proportional to 
$1/(Na)^4$, and the logarithmic $(Na)$-dependence is encoded
in  $\al[1/(Na)]$. Therefore, 
\be
\label{eq:fnnn}
\frac{\pi^2}{36}\,a^4\langle G^2\rangle_{\mathrm{soft}}=
-\frac{1}{N^4}
\sum_{n\geq 0}f_n\al^{n+1}\![1/(Na)]\,,
\ee
and the perturbative expansion of the plaquette on a finite
volume of $N^4$ sites can be written as
\be
\label{OPEpert}
\langle P \rangle_{\mathrm{pert}} (N)=
P_{\mathrm{pert}}(\al)\langle 1 \rangle
+\frac{\pi^2}{36}C_{\G}(\al)\,a^4\langle G^2\rangle_{\mathrm{soft}}
+\mathcal{O}\left(\frac{1}{N^6}\right)\,,
\ee
where 
\be
P_{\mathrm{pert}}(\al)=\sum_{n\geq 0}p_n\al^{n+1}
\ee
and $p_n$ are the infinite volume coefficients that we
are interested in. The constant pre-factor $\pi^2/36$ is chosen
such that the Wilson coefficient, 
which only depends on $\al$, 
is normalized to unity for $\al=0$. It can be expanded in $\al$:
\be
\label{CG}
C_{\G}(\al)=1+\sum_{k\geq 0}c_k\al^{k+1}
\,.
\ee
Since our action is proportional to the plaquette $P$, $C_{\G}$ is
fixed by the conformal trace
anomaly~\cite{DiGiacomo:1990gy,DiGiacomo:1989id}:
\be
\label{CP}
C^{-1}_{\G}(\al)=
-\frac{2\pi\beta(\al)}{\beta_0\al^2}
=1+\frac{\beta_1}{\beta_0}\frac{\al}{4\pi}
+\frac{\beta_2}{\beta_0}\left(\frac{\al}{4\pi}\right)^2
+\frac{\beta_3}{\beta_0}\left(\frac{\al}{4\pi}\right)^3
+\mathcal{O}(\al^4)\,.
\ee
The $\beta$-function coefficients\footnote{We define the $\beta$-function as $\beta(\al)=d\al/d\ln\mu=-\beta_0/(2\pi)\alpha^2-\beta_1/(8\pi^2)\al^3-\cdots$, i.e.\ $\beta_0=11$.} $\beta_j$  are known in the lattice
scheme for $j\leq 3$ (see Eq.~(25) of Ref.~\cite{Bali:2014fea}).

\begin{figure}[t]
\centerline{\includegraphics[width=0.85\textwidth,clip=]{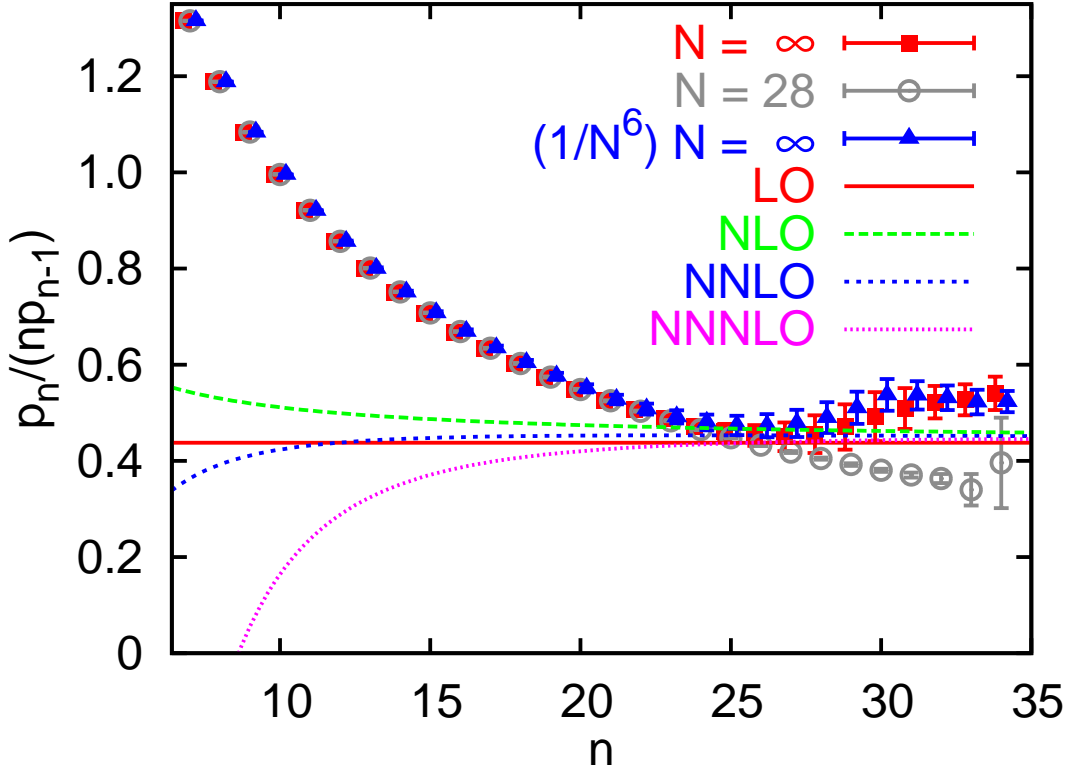}}
\caption{{The ratios
$p_n/(np_{n-1})$ compared with the leading order (LO),
next-to-leading order
(NLO), NNLO and NNNLO predictions of the
$1/n$-expansion \Eqre{th:ratio}.
Only the ``$N=\infty$'' extrapolation includes the systematic uncertainties.
We also show finite volume data for $N=28$, and the result from the
alternative $N\rightarrow\infty$ extrapolation including some $1/N^6$
corrections. The symbols have been shifted slightly horizontally.}
\label{n35}}
\end{figure}

Combining Eqs. (\ref{eq:fnnn}), (\ref{OPEpert}) and (\ref{CG})  gives 
\begin{align}
\label{PpertOPE}
\langle P \rangle_{\mathrm{pert}}(N)
&=\sum_{n\geq 0}\left[p_n-\frac{f_n(N)}{N^4}\right]\al^{n+1}\\\nn
&=
\sum_{n\geq 0}p_n\al^{n+1}
-\frac{1}{N^4}
\left(1+\sum_{k\geq 0}c_k\al^{k+1}(a^{-1})
\right)
\times 
\sum_{n\geq 0}f_n\al^{n+1}((Na)^{-1})
+\mathcal{O}\left(\frac{1}{N^6}\right)\,,
\end{align}
where $f_n(N)$ is a polynomial in powers of $\ln(N)$. 
Fitting this equation to the perturbative lattice results,
the first 35 coefficients $p_n$ were determined
in Ref.~\cite{Bali:2014fea}. The results were confronted
with the expectations from renormalons:
\be
\label{eq:thratio1}
p^{\latt}_n \stackrel{n\rightarrow\infty}{=}
N^{\latt}_{P}\,\left(\frac{\beta_0}{2\pi d}\right)^{\!n}
\frac{\Gamma(n+1+db)}{\Gamma(1+db)} \left[
1+\frac{20.08931\ldots}{n+db}+\frac{505\pm 33}{\left(n+db\right) \left(n+db-1\right)}
+
\mathcal{O}\left(\frac{1}{n^3}\right)
\right]
\,,
\ee
\be
\label{th:ratio}
\frac{p_n}{np_{n-1}}=\frac{\beta_0}{2\pi d}
\left\{1
+\frac{db}{n}
+\frac{db(1-ds_1)}{n^2}
+\frac{db\left[1-3ds_1+d^2b(s_1+2s_2)\right]}{n^3}+
\mathcal{O}\left(\frac{1}{n^4}\right)
\right\}\,,
\ee
where
$b=\beta_1/(2\beta_0^2)$,
$s_1=(\beta_1^2-\beta_0\beta_2)/(4b\beta_0^4)$ and
$s_2=(\beta_1^3-2\beta_0\beta_1\beta_2+\beta_0^2\beta_3)/(16b^2\beta_0^6)$
are defined so that
\be\label{lambdapa}
a=
\frac{1}{\Lambda_{\latt}} \exp\left[-\frac{1}{t}
-b\ln\frac{t}{2}
+s_1 bt-s_2b^2t^2+\cdots\right]
\quad\mathrm{with}\quad t=\frac{\beta_0}{2\pi}\al\,.
\ee
In Fig.~\ref{n35} we compare the infinite volume ratios
$p_n/(np_{n-1})$ to the expectation \Eqre{th:ratio}:
the asymptotic behavior of the perturbative series due to
renormalons is reached around orders $n \sim 27-30$, proving,
for the first time, the existence of the renormalon in the plaquette.
Note that incorporating finite volume effects is compulsory to
see this behavior, since there are no infrared renormalons on a finite lattice.
To parameterize finite size effects we made use of
the purely perturbative OPE Eq.~(\ref{OPEpert}).
The behavior seen in Fig.~\ref{n35}, although computed from
perturbative expansion coefficients, goes beyond the purely perturbative
OPE since it predicts the position of a non-perturbative object
in the Borel plane.

\section{The plaquette: OPE beyond perturbation theory}

Since in NSPT we Taylor expand in powers of $g$ before averaging
over the gauge variables, no  
mass gap is generated.
In non-perturbative Monte-Carlo (MC) lattice simulations an additional 
scale, $\lQ \sim 1/a \, e^{-2\pi/(\beta_0\al)}$,
is generated dynamically (see also Eq.~(\ref{lambdapa})).
However, we can always tune $N$ and $\al$ such that
\be
\label{eq:scalesNP1}
\frac{1}{a} \gg \frac{1}{Na}  \gg \lQ\,.
\ee
In this small-volume situation we encounter a double expansion in powers of 
$a/(Na)$ and $a\lQ$ [or, equivalently,
$(Na)\lQ \times a/(Na)$]. The construction of the OPE
is completely analogous to that of the previous section
and we obtain\footnote{
In the last equality, we approximate the Wilson coefficients by their
perturbative expansions,
neglecting the possibility of non-perturbative contributions
associated to the hard scale $1/a$. These  
would be suppressed by factors $\sim\exp(-2\pi/\alpha)$
and therefore would be sub-leading, relative to the gluon condensate.}
\be
\label{OPEMC}
\langle P\rangle_{\MC} =\frac{1}{Z}\left.\int\![dU_{x,\mu}]\,e^{-S[U]} P[U]
\right|_{\MC}
=
P_{\mathrm{pert}}(\al)\langle 1 \rangle
+\frac{\pi^2}{36}C_{\G}(\al)\,a^4\langle G^2 \rangle_{\MC}
+\mathcal{O}(a^6)\,.
\ee
In the last equality we have factored out the hard scale $1/a$ from the 
scales $1/(Na)$ and $\lQ$, which are encoded in
$\langle G^2\rangle_{\MC}$. Exploiting the right-most
inequality of \Eqre{eq:scalesNP1},
we can expand $\langle G^2 \rangle_{\MC}$ as follows:
\be
\langle G^2\rangle_{\MC}=\langle G^2 \rangle_{\rm soft}\left\{1+\mathcal{O}[\lQ^2 (Na)^2]\right\}
\,.
\ee
Hence, a non-perturbative
small-volume simulation
would yield the same expression as NSPT,
up to non-perturbative corrections that can be made arbitrarily
small by reducing $a$ and therefore
$Na$, keeping $N$ fixed.
In other words, $p_n^{\mathrm{NSPT}}(N)=p_n^{\MC}(N)$ up to
non-perturbative corrections.

We can also consider the limit 
\be
\label{eq:scalesNP2}
\frac{1}{a}  \gg \lQ \gg \frac{1}{Na} \,.
\ee
This is the standard situation realized in non-perturbative lattice simulations.
Again the OPE can be constructed as in the previous section, \Eqre{OPEMC}
holds, and the $p_n$- and $c_n$-values are still the same.
The difference is that now 
\be
\langle G^2 \rangle_{\MC}=
\langle G^2 \rangle_{\NP}\left[1+\mathcal{O}\left(\frac{1}{\lQ^2 (Na)^2}\right)
\right]
\,,
\ee
where $\langle G^2\rangle_{\NP} \sim \lQ^4$ is the so-called
non-perturbative gluon condensate introduced
in Ref.~\cite{Vainshtein:1978wd}. From now on we will call
this quantity simply the ``gluon condensate'' $\langle G^2\rangle$. 
We are now in the position
\begin{itemize}
\item 
to determine the gluon condensate and 
\item
to check the validity of the OPE (at low orders in the $a^2$ scale
expansion) for the case of the plaquette.
\end{itemize}
In order to do so we proceed as follows. 
The perturbative series is divergent due
to renormalons and other, sub-leading, instabilities.\footnote{
The leading renormalon is located at $u=d/2=2$ in the Borel plane,
while the first instanton-anti-instanton contribution occurs at
$u=\beta_0=11N_c/3=11\gg 2$.}
This makes any determination of $\langle G^2 \rangle$ ambiguous,
unless we define precisely how to truncate or how to
approximate the perturbative series. A reasonable definition
that is consistent with
$\langle G^2 \rangle \sim \lQ^4$ can only be given if the asymptotic
behavior of the perturbative series is under control.
This has only been achieved recently~\cite{Bali:2014fea},
where the perturbative expansion of the plaquette was
computed up to $\mathcal{O}(\al^{35})$, see the
previous section. The observed
asymptotic behavior was in full compliance with renormalon
expectations, with successive contributions starting
to diverge for orders around $\al^{27}$--$\al^{30}$ within
the range of couplings $\al$ typically employed in present-day
lattice simulations. 

Extracting the gluon condensate from the
average plaquette was
pioneered in Refs.~\cite{Di Giacomo:1981wt,Kripfganz:1981ri,DiGiacomo:1981dp,Ilgenfritz:1982yx} and
many attempts followed during the next decades,
see, e.g., Refs.~\cite{Alles:1993dn,DiRenzo:1994sy,Ji:1995fe,DiRenzo:1995qc,Burgio:1997hc,Horsley:2001uy,Rakow:2005yn,Meurice:2006cr,Lee:2010hd,Horsley:2012ra}.
These suffered from insufficiently high perturbative orders and,
in some cases, also finite volume
effects. The failure to make contact to the asymptotic regime 
prevented a reliable lattice determination of $\langle G^2\rangle$.
This problem was solved in Ref.~\cite{Bali:2014sja}, which we now summarize.

Truncating the infinite sum at the order
of the minimal contribution provides one definition of the perturbative series.
Varying the truncation order will result in changes of
size $\lQ^4a^4$, where the dimension $d=4$ is fixed
by that of the gluon condensate. We
approximate the asymptotic series by the truncated sum
\be
\label{eq:truncate}
S_P(\al)\equiv S_{n_0}(\al)\,,\quad\mathrm{where}\quad S_n(\al)=\sum_{j=0}^{n}p_j\al^{j+1}\,.
\ee
$n_0\equiv n_0(\al)$ is the order for which $p_{n_0}\al^{{n_0}+1}$ is minimal.
We then obtain the gluon condensate from the relation
\begin{equation}
\langle G^2 \rangle=
\frac{36C_{\G}^{-1}(\al)}{\pi^2a^{4}(\al)}\left[\langle P \rangle_{\MC}(\al)-S_P(\al)\right]+
\mathcal{O}(a^2\lQ^2)
\,.
\label{eq:G2}
\end{equation} 

$C^{-1}_{\G}(\al)$ is proportional to the
$\beta$-function, and the first few terms are known, see Eq.~(\ref{CP}).
The corrections to
$C_{\G}=1$ are small. However, the $\mathcal{O}(\al^2)$ and
$\mathcal{O}(\al^3)$ terms are of similar sizes. We will
account for this uncertainty in our error budget.

Following Eq.~(\ref{eq:G2}), we subtract the
truncated sum $S_P(\al)$ calculated from the coefficients
$p_n$ of Ref.~\cite{Bali:2014fea} from the MC data on $\langle P \rangle_{\MC}(\al)$
of Ref.~\cite{Boyd:1996bx} in the range $\beta \in [5.8,6.65]$ ($\beta=6/g^2$),
where $a(\beta)$ is given by the phenomenological parametrization of 
Ref.~\cite{Necco:2001xg}
($x=\beta-6$)
\be
\label{eq:Necco}
a=r_0\exp\left(-1.6804-1.7331x+0.7849x^2-0.4428x^3\right)\,,
\ee
where $r_0\approx 0.5\,$fm.
This corresponds to
$(a/r_0)^4 \in [3.1\times 10^{-5},5.5\times 10^{-3}]$, covering more than
two orders of magnitude.

\begin{figure}
\includegraphics[width=.47\textwidth,clip=]{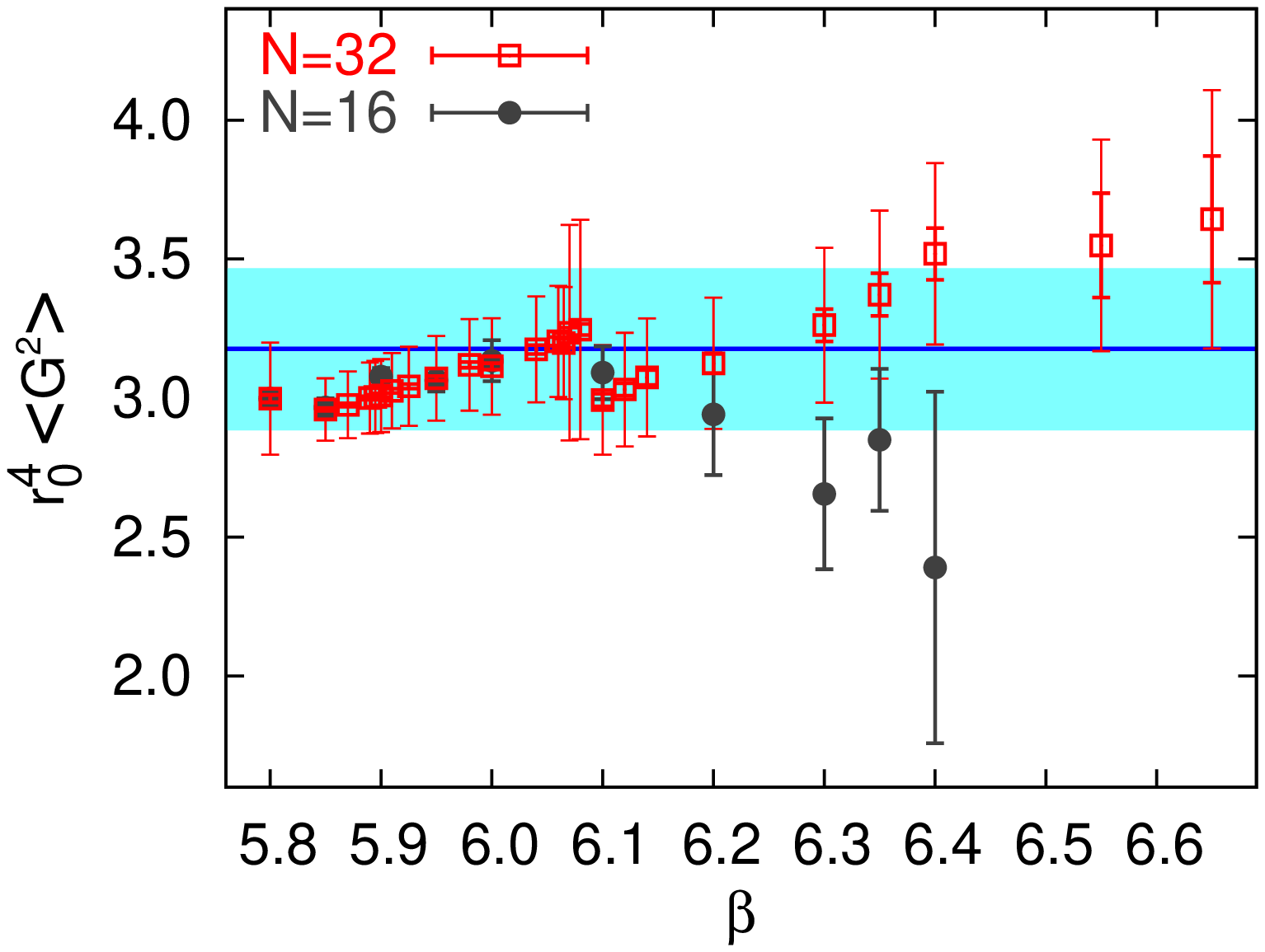}\hspace*{.02\textwidth}
\includegraphics[width=.49\textwidth,clip=]{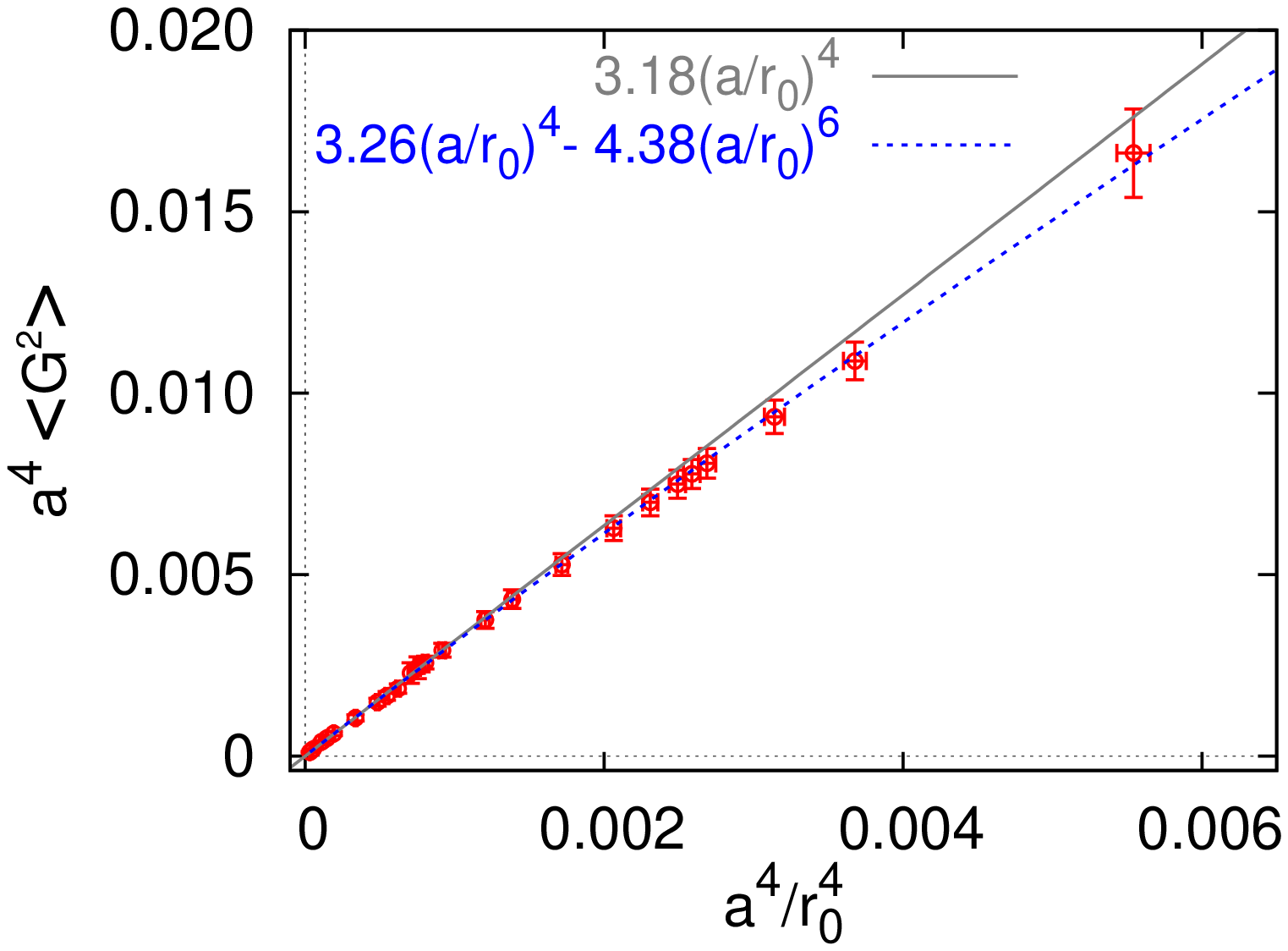}
\caption{\label{fig:PlaqMC}
Left panel: Eq.~(\protect\ref{eq:G2}) evaluated
using the $N=16$ and $N=32$ MC data
of Ref.~\protect\cite{Boyd:1996bx}. The $N=32$ outer error bars include
the error of $S_P(\al)$. The error band is our prediction
for $\langle O_G \rangle$, Eq.~(\protect\ref{eq:G2final}).\newline
Right panel:  Eq.~(\protect\ref{eq:G2}) times $a^4$ vs.\ $a^4(\al)/r_0^4$
from Eq.~(\protect\ref{eq:Necco}). The linear fit
with slope Eq.~(\protect\ref{eq:G2final}) is to the $a^4< 0.0013\,r_0^4$
points only.}
\end{figure}

Multiplying this difference by
$36r_0^4/(\pi^2 C_{\G}a^4)$ gives $r_0^4\langle O_{\G}\rangle$
plus higher order non-perturbative terms. 
We show this combination in the left panel of
Fig.~\ref{fig:PlaqMC}. The smaller error bars represent the
errors of the MC data, the outer error bars (not plotted
for $N=16$) the
total uncertainty, including that of $S_P$. This part of
the error is correlated between different $\beta$-values (see the discussion in Ref~\cite{Bali:2014sja}). 
The MC data
were obtained on volumes $N^4=16^4$ and $N^4=32^4$.
Towards large $\beta$-values the physical volumes
$[Na(\beta)]^4$
will become small, resulting in transitions
into the deconfined phase.
For $\beta<6.3$ we find no significant differences
between the $N=16$ and $N=32$ results. In the analysis
we restrict ourselves to the more precise
$N=32$ data and, to keep finite size effects under control,
to $\beta\leq 6.65$. We also 
limit ourselves to $\beta\geq 5.8$ to avoid large
$\mathcal{O}(a^2)$ corrections. 
At very large $\beta$-values not only does the parametrization Eq.~(\ref{eq:Necco})
break down but obtaining meaningful results becomes challenging
numerically: the individual errors both of $\langle P\rangle_{\MC}(\al)$
and of $S_P(\al)$ somewhat decrease with increasing $\beta$.
However, there are strong cancellations between these two terms, in
particular at large $\beta$-values, since this difference
decreases with $a^{-4}
\sim\Lambda_{\latt}^4\exp(16\pi^2\beta/33)$ on dimensional
grounds while $\langle P\rangle_{\MC}$ depends only logarithmically
on $a$.

The data in the left panel of Fig.~\ref{fig:PlaqMC} show an approximately
constant behavior.\footnote{Note that
$n_0$ increases from 26 to 27 at $\beta=5.85$, from
27 to 28 at $\beta=6.1$ and from 28 to 29
at $\beta=6.55$. This quantization of $n_0$ explains the visible jump
at $\beta=6.1$.} This indicates
that, after subtracting
$S_P(\al)$ from the corresponding MC values
$\langle P \rangle_{\MC}(\al)$, the remainder scales like $a^4$.
This can be seen more explicitly in the right panel
of Fig.~\ref{fig:PlaqMC}, where
we plot this difference in lattice units against $a^4$.
The result is consistent with a linear behavior
but a small curvature seems to be present that can be parametrized as
an $a^6$-correction.
The right-most point ($\beta=5.8$) corresponds to $a^{-1}\simeq
1.45$~GeV while $\beta=6.65$ corresponds to
$a^{-1}\simeq 5.3$~GeV. Note that
$a^2$-terms are clearly ruled out.  

We now determine the gluon condensate.
We obtain the central value and its statistical error
$\langle G^2\rangle=3.177(36)r_0^{-4}$
from averaging the $N=32$ data for $6.0\leq\beta\leq 6.65$.
We now estimate the systematic uncertainties.
Different infinite volume extrapolations of
the $p_n(N)$ data~\cite{Bali:2014fea}
result in changes of the prediction of about $6\%$.
Another $6\%$ error is due to including an $a^6$-term or not
and varying the fit range.
Next there is a scale error of
about 2.5\%, translating $a^4$ into units of $r_0$.
The uncertainty of the perturbatively determined
Wilson coefficient $C_{\G}$ is of a similar size.
This is estimated as the difference between
evaluating Eq.~(\ref{CP}) to $\mathcal{O}(\al^2)$
and to $\mathcal{O}(\al^3)$.
Adding all these sources of uncertainty in quadrature
and using the pure gluodynamics value~\cite{Capitani:1998mq}
$\Lambda_{\MS}=0.602(48)r_0^{-1}$
yields
\be
\label{eq:G2final}
\langle G^2 \rangle=3.18(29)r_0^{-4}=24.2(8.0)\Lambda_{\MS}^4\,.
\ee 

The gluon condensate Eq.~(\ref{eq:GC}) is
independent of the renormalization scale.
However, $\langle G^2\rangle$ was obtained employing one
particular prescription in terms of the observable
and our choice of how to truncate
the perturbative series within a given renormalization scheme.
Different (reasonable) prescriptions can in principle
give different results. One may for instance choose to truncate 
the sum at orders $n_0(\al) \pm \sqrt{n_0(\al)}$ and
the result would still scale like $\lQ^4$. 
We estimated this intrinsic ambiguity of the definition of the
gluon condensate in Ref.~\cite{Bali:2014fea}
as $\delta\langle G^2\rangle=
36/(\pi^2C_{\G}a^4)\,\sqrt{n_0}p_{n_0}\alpha^{n_0+1}$, i.e.\
as $\sqrt{n_0(\al)}$
times the contribution of the minimal term:
\be
\label{eq:prescript}
\delta\langle G^2 \rangle=27(11)\Lambda_{\MS}^4\,.
\ee
Up to $1/n_0$-corrections this definition is scheme- and
scale-independent and corresponds
to the (ambiguous) imaginary part of the Borel integral times
$\sqrt{2/\pi}$.

In QCD with sea quarks
the OPE of the average plaquette or of the Adler function
will receive additional contributions from the chiral condensate.
For instance $\langle G^2\rangle$ needs to be redefined,
adding terms
$\propto\langle \gamma_m(\al)m\bar\psi\psi\rangle$~\cite{Tarrach:1981bi,Bali:2013esa}.
Due to this and the problem of setting a physical scale
in pure gluodynamics, it is difficult to assess the precise
numerical impact
of including sea quarks onto our estimates
\be
\label{eq:G2GeV}
\langle G^2 \rangle\simeq 0.077\, \mathrm{GeV}^4\,,
\quad
\delta\langle G^2 \rangle\simeq 0.087\, \mathrm{GeV}^4\,,
\ee 
which we obtain using $r_0\simeq 0.5$~fm~\cite{Sommer:1993ce}.
While the systematics of applying
Eqs.~(\ref{eq:G2final})--(\ref{eq:prescript})
to full QCD are unknown, our main observations
should still extend to this case.
We remark that our prediction of the gluon condensate
Eq.~(\ref{eq:G2GeV}) is significantly bigger than values
 obtained in one- and two-loop sum rule analyses,
ranging from
0.01~GeV${}^4$~\cite{Vainshtein:1978wd,Ioffe:2002be} up
to 0.02~GeV${}^4$~\cite{Broadhurst:1994qj,Narison:2011xe}.
However, these numbers were not extracted in the
asymptotic regime, which for a $d=4$ renormalon
in the $\MS$ scheme we expect to set in
at orders $n \gtrsim 7$. 
Moreover, we remark that in schemes
without a hard ultraviolet cut-off, like dimensional regularization,
the extraction of $\langle G^2\rangle$ can become obscured by the
possibility of ultraviolet
renormalons. Independent of these considerations,
all these values are smaller than the intrinsic
prescription dependence Eq.~(\ref{eq:prescript}).

\begin{figure}
\includegraphics[width=.8\textwidth,clip=]{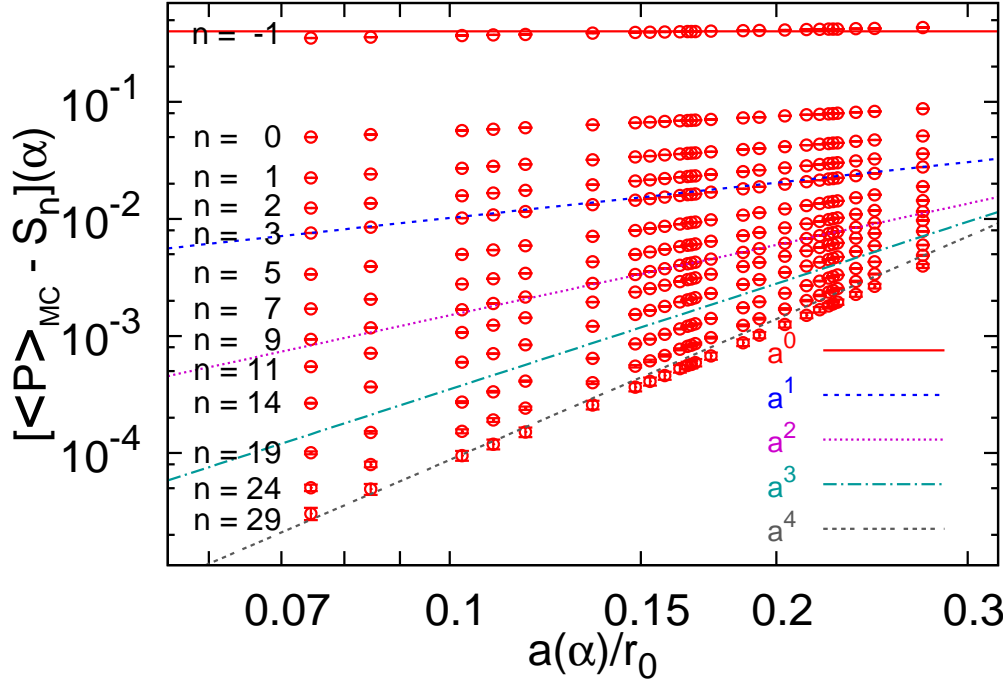}
\caption{Differences $\langle P\rangle_{\MC}(\al)-S_n(\al)$
between MC data and sums truncated
at orders $\al^{n+1}$ ($S_{-1}=0$)
vs.\ $a(\al)/r_0$. The lines $\propto a^j$ are drawn
to guide the eye.
\label{fig:a2}}
\end{figure}

Our analysis confirms the validity of the OPE beyond perturbation theory 
for the case of the plaquette. Our $a^4$-scaling clearly
disfavors suggestions about the existence of
dimension two condensates beyond the standard OPE
framework~\cite{Burgio:1997hc,Chetyrkin:1998yr,Gubarev:2000nz,RuizArriola:2006gq,Andreev:2006vy}.
In fact we can also explain why
an $a^2$-contribution to the plaquette was found in Ref.~\cite{Burgio:1997hc}.
In the log-log plot of Fig.~\ref{fig:a2} we subtract sums $S_n$, truncated at
different fixed orders $\al^{n+1}$, from $\langle P\rangle_{\MC}$.
The scaling continuously moves from $\sim a^0$ at $\mathcal{O}(\al^0)$ to $\sim a^4$ around
$\mathcal{O}(\al^{30})$. Note that truncating at an
$\al$-independent fixed order is inconsistent, explaining why in the figure
we never exactly obtain an $a^4$-slope.
For $n\sim 9$ we reproduce the $a^2$-scaling reported
in Ref.~\cite{Burgio:1997hc} for a fixed order truncation
at $n=7$.
In view of Fig.~\ref{fig:a2}, we conclude that
the observation of this scaling power was accidental.

\section{The binding energy of HQET: OPE in perturbation theory}

The OPE of the plaquette is analogous to the OPE of the vacuum
polarization (or the Adler function) in position space. 
As we already mentioned, the OPE concept of the factorization of scales can
also be applied to more general kinematical settings (in particular
to cases where some scales are defined in Minkowski spacetime).
A prominent example is heavy quark effective theory (HQET). In this case
the term $\langle O_1\rangle$ of \Eqre{eq:ope1}
is replaced by a non-perturbative quantity, the so-called
heavy quark binding energy $\overline{\Lambda}$, that cannot be represented
as an expectation value of a local gauge invariant operator.
We consider the self-energy of a heavy quark in the infinite mass
limit (in other words, the self-energy of a static source in the triplet
representation, for other color representations see Ref.~\cite{Bali:2013pla}),
that is closely related to $\overline{\Lambda}$.
We compute this quantity in close analogy to the case of the plaquette
in lattice regularization.

First we compute the self-energy of the static quark in perturbation theory
(again using NSPT). We obtain this from the Polyakov loop in an
asymmetric volume $N_S^3N_Ta^4$ of
spatial and temporal extents $N_Sa$ and $N_Ta$, respectively:
\be
L(N_S,N_T)=\frac{1}{N_S^3}\sum_{\mathbf{x}}\frac{1}{3}
\mathrm{Tr}\left[\prod_{x_4/a=0}^{N_T-1}U_{x,4}\right]\,,
\ee
or, more specifically, from its logarithm
\begin{equation}
P(N_S,N_T)=-\frac{\ln\langle L(N_S,N_T)\rangle_{\mathrm{pert}}}{a N_T}
=
\sum_{n\geq 0}c_n(N_S,N_T)\al^{n+1}
\,.
\end{equation}
Again, $U_{x,\mu}$ denotes a gauge link and $x=(\mathbf{x},x_4)\in\Lambda_E$
are Euclidean lattice points.
We define the energy of a static source and its perturbative
expansion in a finite spatial volume,
\be
\delta m(N_S)=\lim_{N_T\rightarrow\infty}P(N_S,N_T)=\frac{1}{a}
\sum_{n\geq 0}c_n(N_S)\al^{n+1}\!(1/a)\,,
\quad \mathrm{where} 
\quad
c_n(N_S)=\lim_{N_T\rightarrow\infty}c_n(N_S,N_T)
\,,
\ee
and its infinite volume limit
\be
\label{deltam}
\delta m=\lim_{N_S,N_T\rightarrow\infty}P(N_S,N_T)=\frac{1}{a}\sum_{n= 0}^{\infty}c_n\alpha^{n+1}\!(1/a)\quad\mathrm{with}\quad
c_n=\lim_{N_S\rightarrow\infty}c_n(N_S)
\,.
\ee

We now construct the purely perturbative OPE in a finite volume.
For large $N_S$, we can write [$a^{-1} \gg (N_Sa)^{-1}$]:
\begin{align}
\delta m(N_S)&=\frac{1}{a}\sum_{n\geq 0}^{\infty}\left(c_n-\frac{f_n(N_S)}{N_S}\right)\alpha^{n+1}\!(1/a)+\mathcal{O}{\left(\frac{1}{N_S^2}\right)}\\\nn&=
\frac{1}{a}\sum_{n\geq 0}c_n\alpha^{n+1}\!\!\left(1/a\right)-
\frac{1}{N_Sa}\sum_{n\geq 0}f_n\alpha^{n+1}\!\!\left[1/(N_Sa)\right]
+\mathcal{O}{\left(\frac{1}{N_S^2}\right)}
\,.
\end{align}
Note the similarity between this equation and Eq.~(\ref{PpertOPE}).
$f_n(N_S)=\sum_{j=0}^{n}f_n^{(j)}\ln^j(N_S)$ is again a polynomial
in powers of $\ln(N_S)$, and the $f_n^{(j)}$ are known combinations
of $\beta$-function coefficients and lower order infinite volume
expansion coefficients $c_k$, $k<n$. The main difference with respect to
the gluon condensate is that
the power correction scales like $1/N_S$, 
instead of $1/N_S^4$, and that now the Wilson coefficient is trivial.
This $1/N_S$ scaling also means that the renormalon behavior will show up
at lower orders $n$ of the perturbative expansion. Fitting the
$c_n(N_S)$ data to this equation, the first 20 $c_n$ coefficients
were determined in 
Refs.~\cite{Bauer:2011ws,Bali:2013pla,Bali:2013qla} 
and confronted with the renormalon expectations:\footnote{Here we
deviate from Refs.~\cite{Bauer:2011ws,Bali:2013pla,Bali:2013qla} 
in the definition of the constant $s_2$, see \Eqre{lambdapa}.}
\be
c_n\stackrel{n\rightarrow\infty}{=} N_m\,\left(\frac{\beta_0}{2\pi}\right)^n
\,\frac{\Gamma(n+1+b)}{
\Gamma(1+b)}
\left[
1+\frac{bs_1}{(n+b)}+\frac{b^2\left(s_1^2/2-s_2\right)}{(n+b)(n+b-1)}+ \cdots
\right]\,.
\ee
In the lattice scheme the numerical values
of the above coefficients read $bs_1=1.36095381(11)$ and
$b^2(s_1^2/2-s_2)=5.34(51)$. As expected, the
above expansion converges much faster in $1/n$ than
\Eqre{eq:thratio1}. Calculating the ratio of subsequent
perturbative coefficients gives
\be
\label{eq:ratpol}
\frac{c_{n}}{c_{n-1}}\frac{1}{n} =
\frac{\beta_0}{2\pi}
\left\{1 +\frac{b}{ n} - \frac{b s_1}{n^2}
+\frac{1}{n^3}
\left[2b^2s_2+b(b-1)s_1
\right]
+\mathcal{O}\left(\frac{1}{n^4}\right)
\right\}\,.
\ee
We remark that \Eqre{th:ratio} includes the effect of the
non-trivial Wilson coefficient $C_{\G}$. Therefore, just setting $d=1$ in that equation does not result in Eq.~(\ref{eq:ratpol}) above.

\begin{figure}
\includegraphics[width=0.8\textwidth]{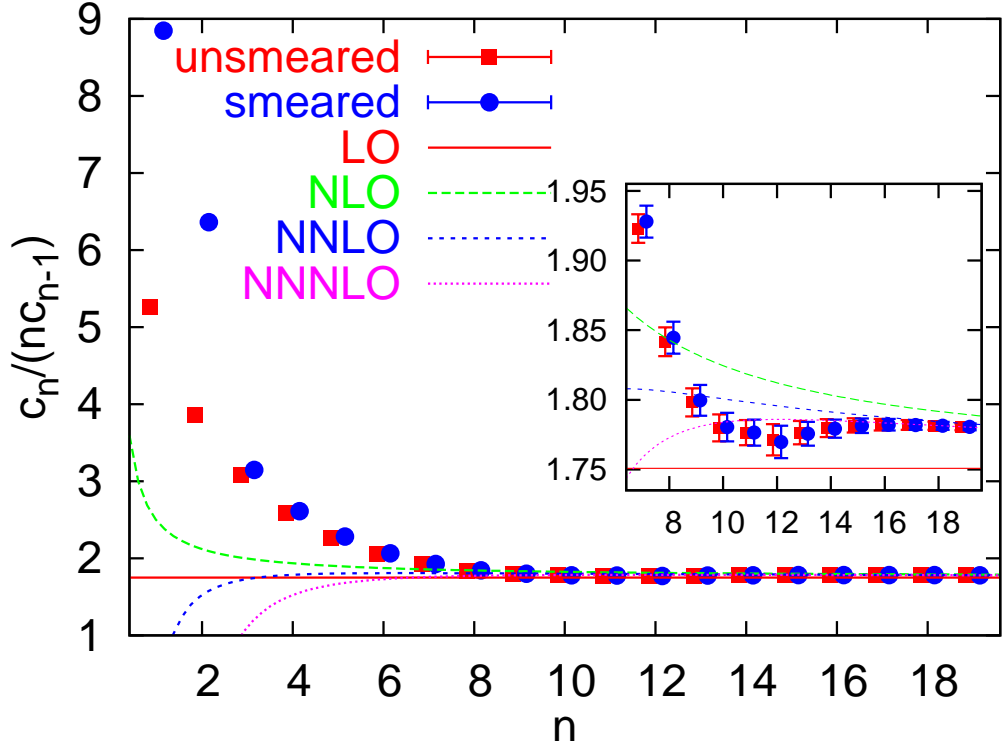}
\caption{Ratios~$c_n/(nc_{n-1})$ of the unsmeared (squares) and smeared
(circles) triplet static self-energy coefficients $c_n$
in comparison to the theoretical prediction
Eq.~(\protect\ref{eq:ratpol}), truncated at different
orders in $1/n$. In the inset we magnify the asymptotic region.
\label{fig:cnratio}
}
\end{figure}

In Fig. \ref{fig:cnratio} we compare the data to \Eqre{eq:ratpol}
for two different lattice discretizations of the covariant
temporal derivative, which amounts to ``smearing'' or not smearing
temporal gauge links. This should not affect 
the infrared behavior and, indeed, beyond the first few orders
the difference becomes invisible. The asymptotic behavior of the perturbative
series due to the renormalon is confirmed in full glory 
for $n \gtrsim 8$, proving the existence of the renormalon behavior
in QCD beyond any reasonable doubt. Again the incorporation of finite volume
effects was decisive to obtain this result.

\section{The binding energy of HQET: OPE beyond perturbation theory}

The methods used for the gluon condensate can also
be applied to other observables.
We now consider the non-perturbative evaluation 
of the energy of a static-light meson on the lattice
and its OPE:
\be
E_{\MC}(\al)=E_{\mathrm{pert}}(\al)+\overline{\Lambda}+\mathcal{O}(a\lQ^2)\,.
\ee
$\overline{\Lambda}$ is the non-perturbative binding energy and $E_{\mathrm{pert}}(\al)=\delta m(\al)$ 
is the self-energy of the static source in perturbation theory, i.e. Eq.~(\ref{deltam}).
In HQET the mass of the $B$ meson is given as
$m_B=m_{b,\mathrm{pert}}+\overline{\Lambda}+\mathcal{O}(1/m_b)$,
where $m_{b,\mathrm{pert}}$ is the $b$-quark pole mass
$m_b^{\mathrm{OS}}$, which suffers from the same renormalon ambiguity as $\delta m$ \cite{Bigi:1994em,Beneke:1994sw,Neubert:1994wq}.

The perturbative expansion of $a\delta m(\al)=\sum_n c_n\al^{n+1}$
was obtained in Refs.~\cite{Bauer:2011ws,Bali:2013pla}
up to $\mathcal{O}(\al^{20})$, see the previous section.
Its intrinsic ambiguity
\be
\label{eq:ambigu}
\delta\overline{\Lambda}=\sqrt{n_0}c_{n_0}\alpha^{n_0+1}=0.748(42)\Lambda_{\MS}=0.450(44)r_0^{-1}
\ee 
was computed in Refs.~\cite{Bali:2013pla,Bali:2013qla}.
MC data for the ground state energy $E_{\MC}$ of a static-light meson
with the Wilson action can be found
in Refs.~\cite{Duncan:1994uq,Allton:1994tt,Ewing:1995ih}. 

\begin{figure}
\includegraphics[width=.8\textwidth,clip=]{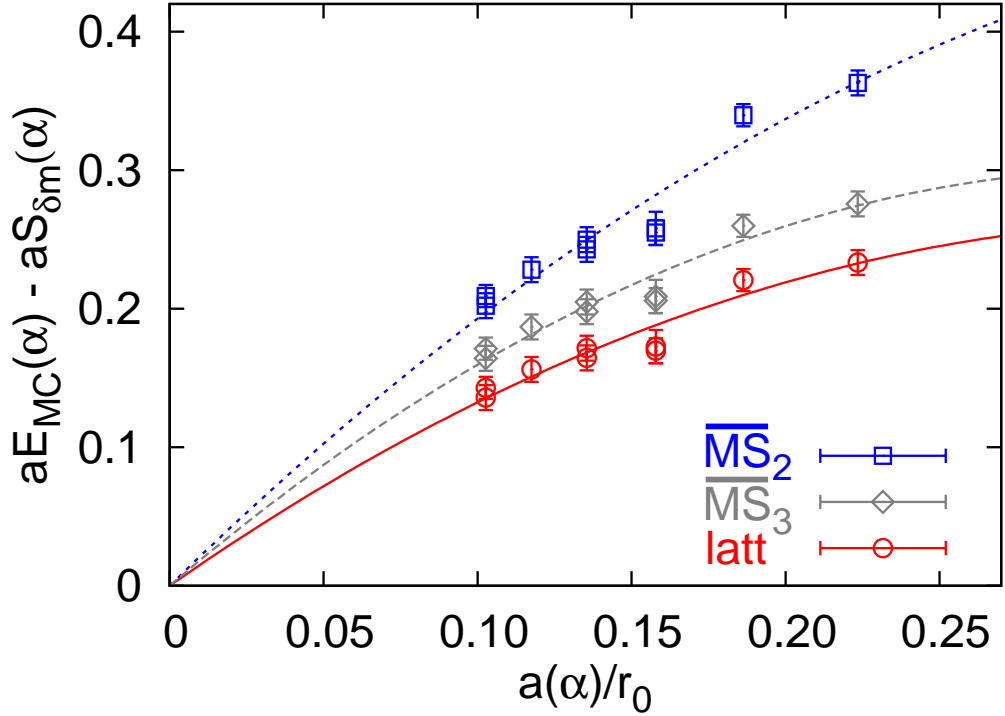}
\caption{$a\overline{\Lambda}=aE_{\MC}-aS_{\delta m}$ vs. $a/r_0$. The
sum $aS_{\delta m}$ Eq.~(\protect\ref{eq:truncsum})
was also converted into the $\MS$ scheme at
two ($\MS_2$) and three ($\MS_3$) loops and truncated at
the respective minimal orders. The curves are fits to
$\overline{\Lambda} a+c a^2/r_0^2$.
\label{fig:lambdabar}}
\end{figure}

Truncating the infinite sum in $\delta m$ at the order
of the minimal contribution provides one definition of the perturbative series. As we did for the plaquette, 
we approximate the asymptotic series by the truncated sum
\be
\label{eq:truncsum}
S_{\delta m}(\al)\equiv \frac{1}{a}\sum_{j=0}^{n_0(\al)}c_j\al^{j+1}\,,
\ee
where
$n_0\equiv n_0(\al)$ is the order for which $c_{n_0}\al^{{n_0}+1}$ is minimal.
While for the
gluon condensate we expected an $a^4$-scaling (see the right
panel of Fig.~\ref{fig:PlaqMC}),  
for $a\overline{\Lambda}=aE_{\MC}-aS_{\delta m}(\al)$ we expect a scaling linear in $a$.
Comforting enough this is what we find, up
to the expected $a\mathcal{O}(a\lQ^2)$ discretization corrections,
see Fig.~\ref{fig:lambdabar}. 
Subtracting the partial sum truncated at orders $n_0(\al)= 6$
from the $\beta\in[5.9,6.4]$
data, we obtain $\overline{\Lambda}= 1.55(8) r_0^{-1}$
from such a linear plus quadratic fit, where
we only give the statistical uncertainty.
The errors of the perturbative coefficients are all
tiny, which allows us to transform
the expansion $a\delta m(\al)$
into $\MS$-like schemes and to compute $\overline{\Lambda}$ accordingly. 
We define the schemes $\MS_2$ and $\MS_3$ by truncating
$\al_{\MS}(a^{-1})=\al(1+d_1\al+d_2\al^2+\ldots)$ exactly at
$\mathcal{O}(\al^3)$ and 
$\mathcal{O}(\al^4)$, respectively. The $d_j$
are known for $j\leq 3$~\cite{Bali:2013pla,Bali:2013qla}.
We typically find $n_0^{\MS_i}(\al_{\MS_i})=2, 3$
and obtain $\overline{\Lambda} \sim 2.17(8) r_0^{-1}$
and $\overline{\Lambda} \sim 1.89(8) r_0^{-1}$,
respectively, see Fig.~\ref{fig:lambdabar}.
We conclude that the changes due to these resummations are
indeed of the size
$\delta\overline{\Lambda}\sim 0.5 r_0^{-1}$,
adding confidence that our definition of the
ambiguity \Eqre{eq:ambigu} is neither a gross overestimate nor an underestimate.
For the plaquette, where we expect $n_0^{\MS} \sim 7$,
we cannot carry out a similar analysis,
due to the extremely high precision that is required
to resolve the differences 
between $S_P(\al)$ and $\langle P \rangle_{\MC}(\al)$, which
largely cancel in Eq.~(\ref{eq:G2}).

\section{Conclusions}

For the first time ever, perturbative expansions at orders
where the asymptotic regime is reached have been obtained and subtracted
from non-perturbative Monte Carlo data of the static-light meson mass
and of the plaquette, thereby validating the OPE for these cases
beyond perturbation
theory. The scaling of the latter difference with the lattice spacing
confirms the dimension $d=4$. Dimension $d<4$ slopes appear only
when subtracting the perturbative series truncated at fixed
pre-asymptotic orders. Therefore, we interpret the lower dimensional ``condensates'' discussed in Ref.~\cite{Chetyrkin:1998yr} as
approximate parametrizations of unaccounted perturbative effects, i.e.\ of the short-distance behavior. These will be observable-dependent,
unlike the non-perturbative gluon condensate. Such simplified
parametrizations introduce unquantifiable errors and, therefore, are of limited phenomenological use. As we have demonstrated above 
(see Fig.~\ref{fig:a2}), even the 
effective dimension of such a ``condensate'' varies when
truncating a perturbative series at different orders. In
Refs.~\cite{Gubarev:2000nz,RuizArriola:2006gq,Andreev:2006vy} various analyses, based on models such as string/gauge duality 
or Regge models, have been made claiming the existence of non-perturbative dimension two condensates. Our results 
strongly suggest that there may be flaws in these derivations.

We observe that $\overline{\Lambda}$ and $\langle G^2\rangle$ do not
depend on the lattice spacing (i.e.\ on the
renormalization scale). In other words, they are renormalization
group invariant quantities, as expected. This is coherent with the
interpretation of these quantities to be of order $\lQ$ and $\lQ^4$,
respectively. However, the values of
$\overline{\Lambda}$ and $\langle G^2\rangle$
will depend on the
details of how the divergent perturbative series is truncated or estimated
(and therefore implicitly also on the scheme used in the perturbative
expansion), as well as on the observable used as an input in the determination.

We have obtained an accurate value of
the gluon condensate in SU(3) gluodynamics, Eq.~(\ref{eq:G2final}).
It is of a similar size as the intrinsic
difference, Eq.~(\ref{eq:prescript}), between
(reasonable) subtraction prescriptions. This result contradicts the 
implicit assumption of sum rule analyses that the renormalon ambiguity is 
much smaller than leading non-perturbative corrections.
The value of the gluon condensate obtained with sum rules
can vary significantly due to this intrinsic
ambiguity if determined using different
prescriptions or truncating at different orders in perturbation theory.
Clearly, the impact of this, e.g., on determinations
of $\alpha_s$ from $\tau$-decays or from lattice simulations needs to be
assessed carefully.

As already mentioned in the previous paragraph,
due to the non-convergent nature of the perturbative series,
the binding energy $\overline{\Lambda}$
and the gluon condensate $\langle G^2\rangle$
are in principle
ill-defined quantities. The intrinsic ambiguity of their definition
can, however, be estimated.
The ambiguity of the HQET binding energy
$\delta \overline{\Lambda}=0.75(4)\Lambda_{\MS}$ as well as the ambiguity
$\delta\langle G^2\rangle=27(11)\Lambda_{\MS}^4$
of the non-perturbative
gluon condensate are scale- and renormalization-scheme independent, at
least up to $1/n_0$-corrections, where $n_0$ is the order of
the minimal term of the perturbative series.
In the first case this ambiguity is significantly
smaller than $\overline{\Lambda}\simeq 2.6\Lambda_{\MS}$.
In the second case the ambiguity is larger
than values typically quoted for $\langle G^2\rangle$,
including our result $\langle G^2\rangle \simeq 24\Lambda_{\MS}^4$. 
The size of $\delta \langle G^2\rangle\gtrsim\langle G^2\rangle$
means that, at least in pure gluodynamics,\footnote{Although our
computations were performed for $N_f=0$, we would not expect the
situation to be qualitatively different including sea quarks.}
$\langle G^2\rangle$ should not be used to estimate the magnitude
of unknown non-perturbative corrections. Instead, 
the ambiguity $\delta \langle G^2\rangle$ should be used
for this purpose. An exception to this rule are situations
where the renormalon ambiguities cancel exactly or are small.
For instance, the perturbative expansion of a difference
between two observables $\langle B_2\rangle-\langle B_1\rangle$
that receive contributions $\propto C_{B_j}\langle G^2\rangle/Q^4$
with a relative normalization such that $C_{B_1}=C_{B_2}+\mathcal{O}(\al)$
will be partially blind to the associated $d=4$ infrared
renormalon if the $\langle B_j\rangle$ are expanded
at the same scale, in the same renormalization scheme and
the same method is used to truncate
both perturbative series.

\begin{theacknowledgments}
This work was supported
in part by the Spanish grants FPA2010-16963 and FPA2011-25948, the Catalan grant
SGR2014-1450 and the German DFG SFB/TRR-55.
\end{theacknowledgments}

\bibliographystyle{aipproc}

\end{document}